\documentclass[10pt,a4paper]{article}
\usepackage[utf8]{inputenc}
\usepackage{amsmath}
\usepackage{amsfonts}
\usepackage{amssymb}
\usepackage{graphicx}
\usepackage{hyperref}
\usepackage{subcaption}
\usepackage{comment}
\usepackage{achemso}
\usepackage{bm}
\usepackage{authblk}
\usepackage{wrapfig}

\title{Cell Dynamic Simulations of Diblock Copolymer/Colloid Systems}
\author[1]{Javier Díaz}
\author[1]{Marco Pinna\thanks{mpinna@lincoln.ac.uk}}
\author[1]{Andrei V. Zvelindovsky}
\author[1]{Adelchi Asta\thanks{ Present adress: Laboratoire PHENIX, Université Pierre et Marie Curie, Case Courrier 51, 4 place Jussieu — 75252 Paris Cedex 5, France    }}
\author[2]{Ignacio Pagonabarraga\thanks{ipagonabarraga@ub.edu}}
\affil[1]{School of Mathematics and Physics, University of Lincoln. Brayford Pool, Lincoln, LN6 7TS, UK}
\affil[2]{Departament de Física de la Matèria Condensada, Universitat de Barcelona, Martí i Franquès 1, 08028 Barcelona, Spain
}

\newcommand{\psici}{\psi_{i}^0}
\newcommand{\rieff}{R^{\text{eff}}_{i}}
\newcommand{\fcal}{ \mathcal{F} }
\newcommand{\rvec}{ \mathbf{r} }
\newcommand{\Rvec}{ \mathbf{R} }

\newcommand{\Fvec}{ \mathbf{F} }

\newcommand{\vvec}{ \mathbf{v} }

\begin{document} \maketitle

\begin{abstract}
The presence of nanoparticles in a diblock copolymer leads to changes in the morphology and properties of the matrix and can produce highly organized hybrid materials. The resulting material properties depend not only on the  polymer composition, but also on the size, shape and surface properties of the colloids. We study the dynamics of this kind of systems using a hybrid mesoscopic approach. A continuum description for the polymer is used, while colloids are individually resolved. The method allows for a variable preference of the colloids, which can have different sizes, to the different components the block copolymer is made of. We can analyze the impact that the nanoparticle preference for either, both or none of the blocks  have on the collective properties of nanoparticle-block copolymer composites. Several experimental results are reproduced covering colloid-induced phase transition, particles' placement within the matrix and the role of incompatibilities between colloids and monomers. 
\end{abstract}
\begin{figure}[t]
\centering
\includegraphics[width=0.5\linewidth]{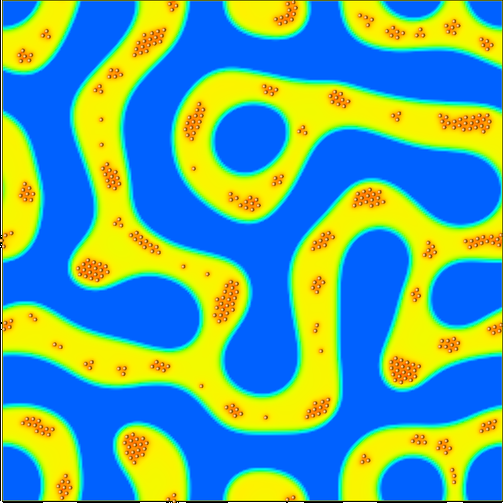}
\end{figure}

\clearpage
\section{Introduction}
The addition of nanoparticles (NP's) to Block Copolymers (BCP) results in a nanomaterial which properties are significantly different from the one of purely one of its components (e.g. changes in the BCP morphology\cite{Bates1999})   
  The resulting hybrid material can be used as a catalyst in separation processes or as photonic-gap materials\cite{Templin_1997,Zhao_1998,Chan_1999} . The nonlinear optical properties of BCP are enhanced with the inclusion of selective NP's \cite{Tsuchiya} . Following the work done in our previous  article \cite{macro_Pago} we study colloidal particles of the size of the order of magnitude of the BCP domain which in our case is in the range of a few nanometers. In this regime,both NP's and BCP evolve in a similar time scale.
 
 Contrary to previous works (e.g. Thompson \textit{et al}\cite{Thompson2001} ) in which the internal structure of the BCP chain was taken into account, we will take a mesoscopic perspective which allows us to perform longer simulations over a larger length scale. To this end we describe the polymer chains through their monomer concentration leading to a free energy that describes both the polymer/polymer interaction as well as the interaction between the polymer chains and the colloidal inclusions. The computational model is hybrid since we combine a polymer description where the monomer concentration is treated as a continuum while we resolve the individual dynamics of the colloidal particles, assuming it is described by Brownian dynamics. This mesocopic approach  puts forward a simplified model that will allow for computationally inexpensive simulations.

Previous works have studied systems of  polymers melts including particles, using different techniques such as strong-segregation theory\cite{Kim2002} , Monte-Carlo\cite{Wang2003,Huh2000,Kang2008} , SFCT combined with DFFT\cite{Sides2006,Thompson2001} , molecular dynamics\cite{Schultz_2005} , dissipative particle dynamics\cite{Maly2008,Posocco_2010}  as well as methods than combine a phenomenological Cahn-Hillard picture  for the fluid with Brownian dynamics for the particles\cite{Balazs_2000,Ginzburg2000,Ginzburg2002} .   

Here we report on further extension of our previous work \cite{macro_Pago} which allow to simulate two different kinds of particles, which may differ in their size as well as their affinity. We aim to improve our understanding of the role of NP's in a BCP matrix and their assembly and placement within their soluble block. 

While the previously mentioned works studied similar systems, the complexity of the simulations of these hybrid systems results in computationally expensive numerical calculations. From a mesoscopic point of view we can reduce the complexity, thus allowing us to perform a considerable number of simulations exploring a particular parameter. The volume fraction of NP's and their affinity will be carefully studied in order to analyse its role in the morphology of the BCP as well as the assembly and placement of the NP's within the matrix.

\section{Model}

The evolution of the BCP/colloids  system is determined by the excess free energy which can be separated as 
\begin{equation}
\fcal _{tot} = \fcal_{pol}+\fcal_{cc} +\fcal_{cpl}
\end{equation}
with $\fcal_{pol}$ being the free energy functional of the BCP melt, $\fcal_{cc}$ the colloid-colloid interaction and the last contribution being the coupling term between the block copolymer and the colloids. 

\subsection{Polymer Dynamics: Cell Dynamics Simulations}

The diblock copolymer is characterized by the order parameter $\psi ( \rvec ,t   )$ which represents the differences in the local volume fraction for the copolymer A and B 
\begin{equation}
\psi (\rvec,t )= 
\phi_A (\rvec,t)
-
\phi_B (\rvec,t)
+(1-2f)
\end{equation}
with respect to the relative volume fraction of A monomers in the diblock, $f= N_A/ (N_A +N_B)$.

The order parameter must follow the continuity equation in order to satisfy the mass conservation of the polymer: 
\begin{equation}
\frac{\partial\psi ( \rvec, t )}{\partial t}=
-\nabla\cdot \mathbf{j} (\rvec ,t ) 
\end{equation} 

If the polymer relaxes diffusely towards equilibrium, the order parameter flux can be expressed in the form 
\begin{equation}
\mathbf{j } (\rvec,t    )=
-M \ \nabla \mu (\rvec , t )
\end{equation}
as a linear function of the order chemical potential
\begin{equation}
\mu (\rvec , t )=
\frac{\delta \fcal_{tot}  [ \psi] }{ \delta \psi}
\end{equation}

Introducing these equations into the continuity equation and taking into account the thermal fluctuations we obtain the Cahn-Hilliard-Cook equation (CHC)
\begin{equation}
\frac{\partial\psi ( \rvec, t )}{\partial t}=
M\ \nabla^2 \left[
\frac{\delta \fcal_{tot}  [ \psi] }{ \delta \psi}
+
\xi ( \rvec, t)
\right]
\label{eq:cahn}
\end{equation}
where $M$ is a phenomenological  mobility constant and $\xi$ is a white Gaussian random noise which satisfies the fluctuation-dissipation theorem\cite{Ball_1990} . 

The copolymer free energy is a functional of the local order parameter which can be expressed in terms of the thermal energy $k_B T$ as
\begin{equation}
\begin{split}
\fcal_{pol}  [ \psi (\rvec ) ]=
\int d\rvec \left[
H(\psi) +\frac{1}{2} D | \nabla\psi  |^2   
\right]
+\\
\frac{1}{2} B \int d\rvec  \int d\rvec' \ 
G(\rvec -\rvec ' )\psi(\rvec)\psi(\rvec') 
 \end{split}
\end{equation}
where the first and second terms are the short and the long-range interaction terms respectively, the coefficient $D$ is a positive constant that accounts for the cost of local polymer concentration inhomogeneities, the Green function $G(\rvec-\rvec' )$ for the laplace Equation satisfies $\nabla^2 G(\rvec-\rvec') = -\delta (\rvec-\rvec')$, $B$  is a parameter that introduces a chain-length dependence to the free energy\cite{Hamley_2000} and $H (\psi)$ is the local free energy \cite{Hamley_2000,Ren_2001_Macr} , 
\begin{equation}
\begin{split}
 H(\psi )  = 
 \frac{1}{2}\left[   
 -\tau+ A(1-2f)^2
 \right]   \psi ^2 \\
 +\frac{1}{3} v (1-2f)\psi^3 
 +\frac{1}{4} \psi^4
 \end{split}
 \label{eq:Hpsi}
\end{equation}
where $\tau,A,v,u $ are phenomenological parameters\cite{Ren_2001_Macr} which can be related to the block-copolymer molecular specificity. Previous works\cite{macro_Pago,Ren_2001_Macr,Ohta_1986} describe the connection of these effective parameters to the BCP molecular composition.  $\tau ' = -\tau+A(1-2f)^2$, $D$ and $B$ can be expressed\cite{Ohta_1986} in terms of degree of polymerization $N$, the segment length $b$  and the Flory-Huggins parameter $\chi$(inversely proportional to temperature) as 
\begin{equation}
\tau' =-\frac{1}{2N}\left[  N\chi -\frac{s(f)}{4f^2 (1-f^2}   \right];\ 
D=\frac{b^2}{48f(1-f)}; \ B=\frac{9}{4N^2b^2f^2(1-f)^2}
\end{equation}
 $\tau'$ accounts for the net atractive/repulsive interaction between monomers. $s(f)$ is an empirical fitting function of the order of $1$ for the range of values of $f$ \cite{Ohta_1986,Pinna_2008_Pol} . Parameters $D$ and $B$ -respectively governing lamella interface thickness and domain size- are written in dimensionless form by defining $\tilde{D}=D/a_0^2$ and $\tilde{B}=B\ a_0^2$ (for simplicity we will drop this notation), $a_0$ being the lattice spacing. Subsequently, we will consider $u$ and $v$  constants\cite{Leibler_1980}, which define all the parameters identifying the BCP local  free energy $H(\psi)$ . As  previously shown \cite{Sevink_2011,Pinna_2009} , CDS can be used along with more detailed approaches like dynamics self-consistent field theory (DSCFT), using CDS as a precursor in exploring parameter space due to the computationally inexpensiveness nature of CDS.

We can express the time evolution of $\psi$ , Equation \ref{eq:cahn}, using CDS as 
\begin{equation}  
\begin{split}
\psi ( \rvec_i , t+1   )= \psi (\rvec_i,t )-  
\delta t [ 
<< \Gamma (\rvec_i, t )>> \\
- \Gamma (\rvec_i, t ) +  
B     [ 1- P (\rvec_i, t) \psi (\rvec_i,t )]   -\eta \xi (\rvec_i,  t)       ] 
]
\end{split}
\label{eq:time_evol}
\end{equation}
$\rvec_i$ being the position of the node $i$ at a time $t\delta t$, and the isotropic discrete  laplacian for a quantity $X$ is given by \cite{Oono_1988}
$\frac{1}{a_0^2}  [ << X >> -X  ] $. Specifically, we  will use 
\begin{equation}
<<\psi >> = \frac{1}{6}  \sum_{NN}  \psi   +\frac{1}{12} \sum _{NNN} \psi
\end{equation} 
NN, NNN meaning nearest neighbors, next-nearest neighbors, and next-next-nearest neighbors, respectively. 

In Equation \ref{eq:time_evol} we have introduced the auxiliary function 
\begin{equation}
\begin{split}
\Gamma (\rvec, t ) =
g( \psi (\rvec, t)  )- \psi (\rvec, t)+\\
D \left[ 
<<   \psi (\rvec, t)    >>   -\psi (\rvec, t)
\right]
\end{split}
\end{equation}
and also, the map function \cite{Bahiana_1990,Ren_2001_Macr}
\begin{equation}
g (\psi)= -\tau ' \psi -v (1-2f)\psi^2 -u \psi^3
\end{equation}

Contrary to previous CDS models \cite{Feng_2004,Ren_2002} , in Equation \ref{eq:time_evol} we have introduced a function $P (\rvec, t )$ that takes into account the volume excluded by the colloid particle, with values $P=0$ in a unbound medium and $P=1$ inside a solid particle.  

\subsection{Colloid Dynamics: Brownian Motion } 

\begin{figure}[h]
\centering
\includegraphics[width=0.5\linewidth]{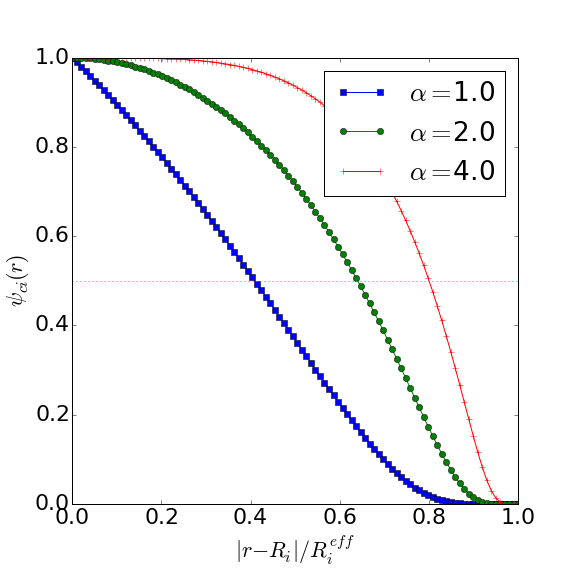}
\caption{Tagged field $\psi_{ci}$ of one particle for several values of $\alpha$. The horizontal line cuts $\psi_{ci}$ at the distance we consider the hard-core radius of the colloid $R_i^0$ . }
\label{fig:model.psi}
\end{figure}

In addition to the described continuous description of the polymer melt we want to model the individual motion of a suspension of $N$ particles in a block copolymer. In order to facilitate the polymer/colloid coupling description, we  assign a continuous field that represents the particles\cite{Tanaka_2000} :  an undeformable tagged field $\psi_{c,i}$ for a particle $i$ centered around the position $\Rvec_i$ (not restricted to lattice nodes) that changes continuously in space while moving rigidly around particle $i$. For computer efficiency it is convenient to introduce a field that is compact. In addition, in order to avoid non-finite values for the derivative of $\psi _{c,i}$, we select a form 
\begin{equation}
\psi_{c,i} = 
exp\left[
1-\frac{1}{1-\left( \frac{| \rvec -\Rvec_i  |}{\rieff}  \right)^\alpha} 
\right]
\label{eq:psici}
\end{equation}
which allows for a cutoff distance at $\rieff$, which is defined in terms of the hard-core radius as $\rieff = R_i^0  \left( 1+1/ln 2    \right)^{1/\alpha} $, meaning that the tagged field has decreased to $1/2$ at $R_0^i$. The parameter $\alpha$ can be tuned to modify the sharpness of the field as we can see in Figure  \ref{fig:model.psi}. Continuity of both $\psici$ and its derivative holds  if we use Equation \ref{eq:psici}.

Summarizing, the overall colloidal suspension is described through the superposition of all the contributions by each particles as in 
\begin{equation}
\psi (\rvec , t  )=
\sum_{i=1}^{N}  \psi_{c,i} (\rvec , t  )
\end{equation}

For the colloid/colloid interaction we propose a pairwise interaction derived from a Yukawa potential, 
\begin{equation}
\begin{split}
V  (\Rvec_i , \Rvec_j      )= \\
U_0   \left\lbrace
\frac{exp\left[  -\alpha_0 \left(  \frac{R_{ij}}{R_{ij}^0}  -1  \right)     \right] }{\frac{R_{ij}}{R_{ij}^0} +\beta_0}- 
\frac{e^{-\eta }}{ \frac{ \eta}{\alpha_0} +1+\beta_0}
\right\rbrace
\end{split}
\end{equation}
where $R_{ij}=| \Rvec_i- \Rvec_j        |$ is the distance between the center of mass of the colloids and $R_{ij}^0=R_i^0  +R_j^0 $ is the distance between two particles when they touch as hard spheres. Parameter $\alpha_0$ determines the steepness of the potential while $\beta_0$ prevents the potential from divergence for complete overlap. A cutoff is introduced for computer efficiency at $R^*_c$, written in terms of $\eta $ as 
\begin{equation}
\frac{R^*_c}{R_{ij}^0}=1+\frac{\eta}{\alpha_0}
\end{equation}

Given this potential the pairwise colloidal force can be derived as $\Fvec^{cc} _i =- \frac{\partial}{\partial \Rvec _i} \fcal_{cc} $ and thus the Equation of motion in a Langevin dynamics scheme would be 
\begin{equation}
m_i  \frac{d \vvec_i }{dt} =
 -\gamma_i \vvec_i + \Fvec^{tot}_i + \sqrt{2k_B T \gamma_i}\bm{\xi}
\end{equation}
with $m_i$ as the particle's mass, $\Fvec^{tot}_i$ is the total force done both by the surrounding polymer as by the rest of the colloids, while $\gamma_i$ stands for the friction exerted by the polymer melt. The thermal motion of the colloids is determined by the temperature through the random term $ \bm{\xi}_i$, with correlations $ <\xi_i (t) >=0  $  and  $  <\xi_i (t), \xi_j (t')>    = \delta_{ij} \delta (t-t')        $ . If we neglect inertia, the Langevin equation reduces, in this overdamped regime, to the Brownian dynamic equation that governs the colloidal dynamics 
\begin{equation}
\vvec _i = \frac{1}{\gamma_i	} \left( \Fvec^{cc}_i+ \Fvec^{cpl}_i +  \sqrt{2k_B T \gamma_i}  \bm{\xi} \right)
\end{equation}

\subsection{Polymer/Colloid Interaction }

The interaction between the polymer and colloids is included through a contribution to the free energy $\fcal_{cpl}$, which must take into account the fact that colloids may have a preference for one of the components of the block copolymer. The simplest free energy that satisfies that is 
\begin{equation}
\fcal_{cpl} =
\sum _{i=1}^N 
\sigma \int d \rvec\ \psi_{c,i} (\rvec) \left[ \psi (\rvec )-\psi_0^i   \right]^2
\label{eq:coupling}
\end{equation}
where $\sigma$ defines the strength of the interaction between polymer and colloids, and $\psi_0^i$ describes the affinity of particle $i$ with one of the copolymer blocks. The value of $\psi_0^i$ stands for the preferred order parameter value for the NP, thus minimising the coupling free energy. $\psi_0^i=0$ will result in neutral NP's while positive or negative values of the affinity means a preference for one of the blocks in the BCP. The affinity is associated with the chemical compatibility  of the NP with the monomers. 

Associated with this term in the free energy there is a corresponding additional force acting on each colloid. For  particle $i$, the force felt due to the nearby polymer inhomogeneities is 
\begin{equation}
\begin{split}
\mathbf{F}^{cpl}_i=
-\sigma \int d\rvec  \ \left[ \psi (\rvec) -\psi_0^i      \right]^2  \\
   \frac{\psi_{c,i} (\rvec) }{  \left[ 1- \left(  \frac{|\rvec- \Rvec_i |}{\rieff}  \right)^\alpha      \right]^2 }  \frac{\alpha}{ {\rieff}^\alpha} | \rvec -\Rvec_i |^{\alpha -1} \frac{\rvec -\Rvec_i }{ \rvec -\Rvec_i |} 
\end{split}   
\end{equation}
and the corresponding term in the copolymer chemical potential that colloids induce in the copolymer dynamics is 
\begin{equation}
\mu_{cpl} =\frac{\delta \fcal_{cpl}}{\delta \psi} = 
2\sigma \sum_{i=1}^{N}\ \psi_{c,i}(\rvec)\left[ \psi (\rvec) -\psi_0^i      \right]
\end{equation}

As opposed to previous approaches\cite{Balazs_2000,Ginzburg2000,Ginzburg2002} , we treat individual colloids through a soft interaction which allows for a simpler computational treatment and a more straight-forward interpretation of the interactions both between colloids as with the copolymer medium. 

Two characteristic time scales appear in our model. BCP diffusive time scale is given by\cite{Guo_2007} $a_0^2/M$ while the colloidal diffusion constant $D=k_B T /\gamma$ sets the NP's diffusive time scale as $R_0^2/D$.

\section{Results}

In this section we present the results of the simulations of our model. Unless otherwise specified, we will set the polymer and colloid parameters to $A=1,\ D=0.5,\ v=1.5,\ u=0.5,\ \tau=0.3,\  k_BT =1,\ \alpha=2,\ \eta=0,\ \beta_0 =0 ,\ \alpha_0 =1 \text{and}\ U_0 =\sigma = 1$. $B=0.01$, except in the last section in which $B=0.001$ in order to tune the domain's size.  The total A-monomer ratio is fixed to $f=0.5$ for the symmetric lamella case, while $f=0.4$ for cylindrical morphologies.  Particle's friction constant is $\gamma= 25.0$, except for the last subsection in which $\gamma=10.0$ accordingly to colloid's radius. The grid size is defined as $a_0=1/4$, except for the hexagonal packing simulations in which we used $a_0=1/\sqrt{7}$ in order to have smaller BCP domains. The time step is changed accordingly as $\delta t =0.05$ and $\delta t =0.2$, while $M=0.1$ for all our simulations. We disregard random fluxes in the BCP setting $\eta = 0$ , for simplicity's sake. 

Unless otherwise specified we will analyze our model on a 2D system of size $256\times 256$ grid spacings. We will explore the collective behavior of BCP/colloid composites changing the colloidal area fraction. From this value one can infer the number of colloidal particles used in different simulations. The volume fraction is calculated using the effective radius, rather than the hard core one. Contrary to that, figures show hard-core radius, as can be appreciated from the coating of BCP surrounding the hard-core. 

Initially the order parameter is an uniformly random distribution while for the colloids we choose an initial non-overlapping random configuration.  

In this section we will apply our model to study the effects that colloidal particles induce in the morphology of a BCP depending on the chemical properties of the NP's surface,  which is described in our model as the affinity towards a value of order parameter. The role of size, chemical composition and compatibility of NP's will be analyzed with regard to its effect on  the colloidal assembly in the BCP matrix. 
\subsection{Phase transition induced by colloids}

The phase diagram of a purely A-B diblock copolymer system is well known \cite{Matsen_PRL_1994}  and for the most simple case we can differentiate between lamellar and clylindrical phases depending on whether the total concentration of A and B monomers is  symmetrical or not. For the second case the shortest block (referred here as the minority phase) forms domains which are hexagonally organized. In this subsection we analyze the ability of the colloids to distort the underlying BCP morphology, promoting a phase transition as we vary both their concentration and affinity

Several theoretical and experimental works have come across the change in the morphology of the BCP induced by the presence of colloids. In this section we apply our model in an attempt to reproduce these results. In contrast with previous works, our model also covers the time evolution towards the equilibrium and is able to reach larger system sizes.

\subsubsection{From Lamellae to Cylinders}

We start analyzing  the effect of relatively large ($ R_i^0 =2.5$) colloids that have a strong affinity towards one of the blocks ($\psici = 0.5$).

\begin{figure}[h]
\centering
\includegraphics[width=1.0\textwidth]{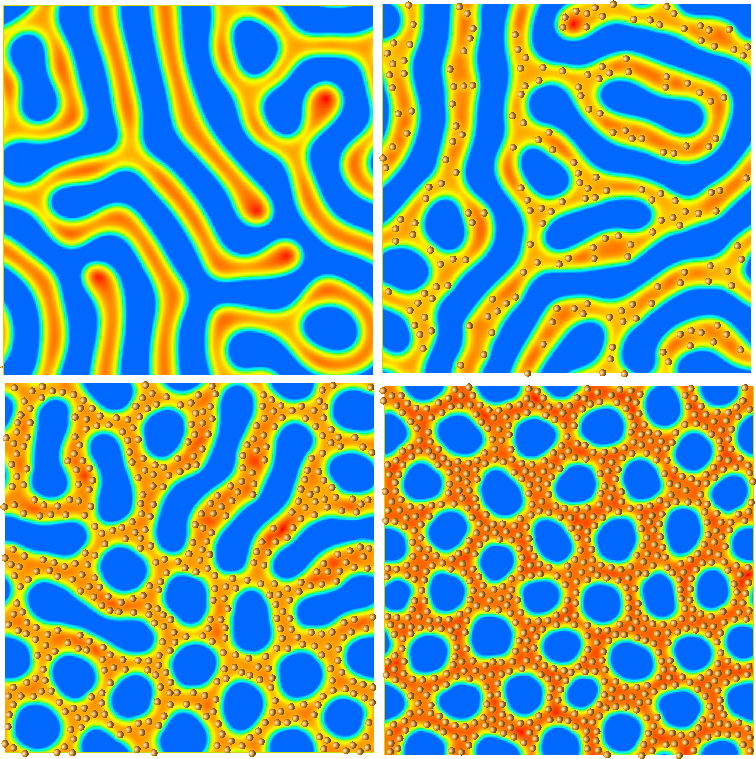}
\caption{Phase transition-lamella to ordered cylindrical-  induced by NP's as we increase their volume fraction with values $\rho=0 \%, 15\%, 35\%,55\% $ for top-left, top-right, bottom-left and bottom-right respectively.}
    \label{fig:l2c}
\end{figure}

Figure \ref{fig:l2c}  illustrates the morphological changes induced in a  BCP lamellar matrix as the NP concentration increases . Figure \ref{fig:l2c} top-left shows a typical lamella morphology without colloids. In the low density regime (Figure \ref{fig:l2c} top-right, $\rho= 15 \%$) the domains are shorter and drop-like domains appear as a result of colloids breaking lamella domains. The transition is not complete in Figure \ref{fig:l2c} bottom-left for $\rho = 35 \%$ but coexistence of lamella and cylindrical phase is observed. At larger colloidal concentration, NPs are closely packed in the BCP domain they are more miscible in, and favor a distortion of the BCP domains, inducing a phase transition. This process is due to the influence of colloids in their preferred copolymer, which effectively increases the fraction of the domain in which they are soluble. Our model reflects the mechanism by which one type of copolymer is excluded from the volume occupied by the NP. This process of phase transition from lamella to ordered cylindrical morphology is in agreement with the experimental results by Kim \textit{et al} \cite{kim2005nanoparticle} and Halevi \textit{et al} \cite{halevi2014} for which our model reproduces particularly well the coexistence of lamella and cylindrical domains, as shown in Figure \ref{fig:l2c} top-right.

\begin{figure}[h]
\centering
\includegraphics[width=1.0\textwidth]{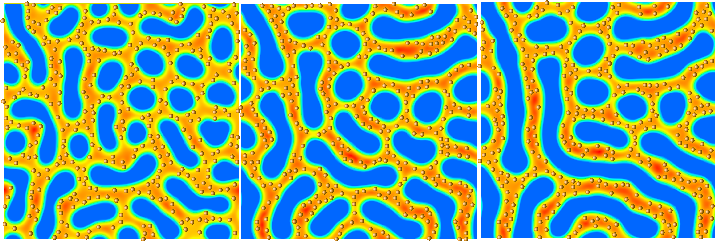}
\caption{ Time evolution of a lamella BCP/colloid with volume fraction  $\rho = 25 \%$ for different stage of the simulation. From left to right: $t= 0.5 \times 10^6 $ steps, $t= 5.0 \times 10^6 $ steps and $t= 14.2 \times 10^6 $ steps.  }
    \label{fig:l2c.time}
\end{figure}

 Intermediate stages of the phase transition kinetics exhibit interesting features.  Figure \ref{fig:l2c.time} shows snapshots of the time evolution. Initially, small cylindrical domains are formed. These domains can coalesce leading, at longer times, to larger cylindrical domains that coexist with  long lamella domains formed by cylinder coalescence.

\begin{figure}[h]
\centering
\includegraphics[width=1.0\linewidth]{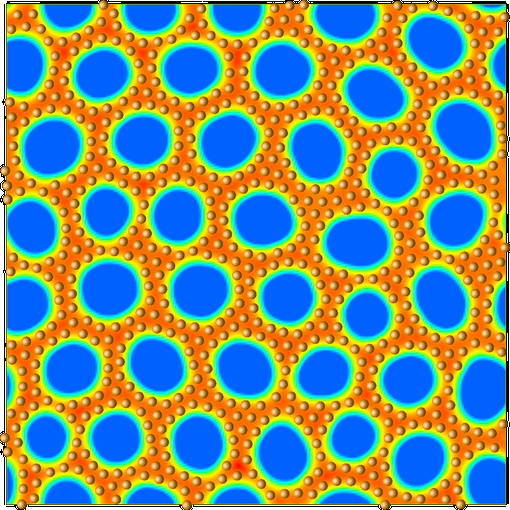}
\caption{A system with a similarly high volume fraction of NP's as in Figure \ref{fig:l2c} bottom-right  but with a temperature $T=0.1$ for colloids.    }
\label{fig:phase.l2c.50.t01}
\end{figure}

For very high filling fraction, the particles attempt to minimize the occupied area by packing closely. This leads to an hexagonal packing if we decrease the colloidal temperature, thus reducing the random component of the colloid's motion. Figure \ref{fig:phase.l2c.50.t01} shows the same system as in Figure \ref{fig:l2c} bottom-right with an slightly lower number of particles $\rho=50.0 \%$ and $T=0.1$ reducing the thermal component of the colloidal velocity.  This phenomena will appear again when we reach high concentrations and can be regarded as a transition from liquid to solid phase for the colloids, guided by the interplay between temperature and  concentration. Further analysis is needed to obtain a clearer picture of the coupling between NP and BCP phase transitions.

\subsubsection{From Cylinders to Lamellae}
 
\begin{figure}[h]
\centering
\includegraphics[width=0.8\textwidth]{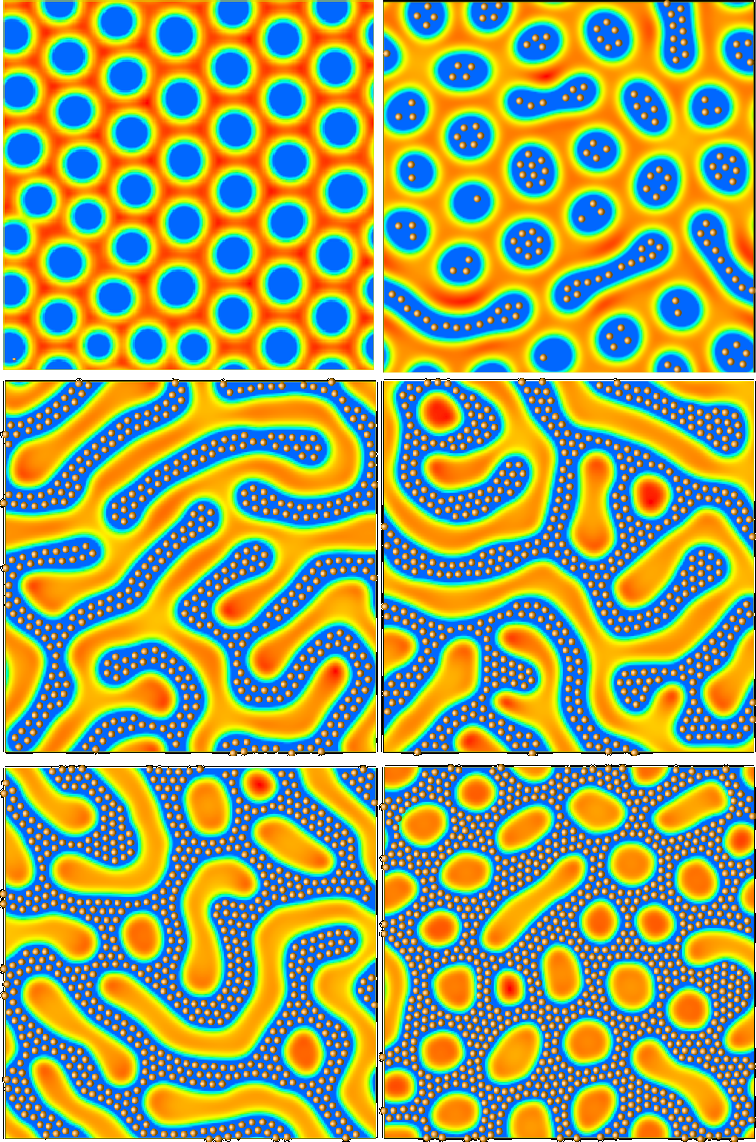}
    \caption{ Phase transition-cylindrical to lamella,then to disordered cylindrical-  induced by NP's as we increase their volume fraction with values $\rho = 0.0 \%,12.0 \%,32.0 \%,38.0 \%,46.8 \%,65.0 \%,$ from top-left to bottom-right, respectively. }
    \label{fig:c2l}
    \end{figure}

Contrary to the previous process, if we now begin with a cylindrical BCP morphology, such as in Figure   \ref{fig:c2l} top-left, and we increase the particle concentration ($R_0=2.5$ ) with a strong affinity towards the minority phase ($\psici= - 0.8$) and $T=0.1$, the effective total concentration of the minority phase is increased. In Figure \ref{fig:c2l} top-right this process is clear as colloids distort the drop-like domains due to the repulsive interaction between particles leading to elongated domains  up to a point in which phase transition is achieved and BCP morphology  changes to lamella (Figure \ref{fig:c2l} center-left). For an even further value of the filling fraction, closely packed NP's increase their occupied area. It triggers an additional phase transition from lamella to  disordered cylindrical (the previously minority phase being the dominant now) which can be seen in different stages for an increasing number of NP's in Figures \ref{fig:c2l} cente-right and bottom , in which drop-like domains become the majority.  


\begin{figure}[h]
\centering
\includegraphics[width=\textwidth]{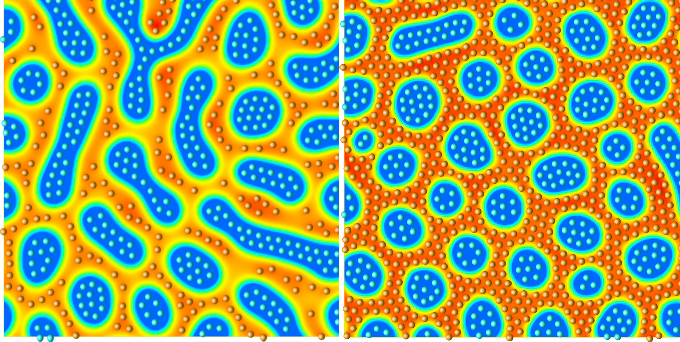}
\caption{Effect of the inclusion of NP's with opposite affinity to a fixed number of colloids with affinity towards the center of the drop-like domains. On the left there are a colloids in the yellow phase with volume fraction $\rho_2= 15 \%$ and $\rho_2 = 50 \%$ on the right picture. }
\label{fig:phase.prevent}
\end{figure}

The phase transition from cylindrical to lamella can be prevented if we include colloids with the opposite affinity ($\psici = +0.5$). This situation can be simulated using the same BCP morphology as in the previous study and a fixed number of colloids with radius $R=2.5$ and volume fraction  $\rho_0= 22\%$  with affinity towards the minority phase,  while changing the number of identically sized NP's in the other BCP domain.  Figure \ref{fig:phase.prevent} shows this effect as in the left picture a low volume fraction of colloids in the majority phase ($\rho_2= 15 \%$) allows the formation of a majority of  lamella domains, while in the right hand side Figure the growth in the effective volume fraction of the majority copolymer due to a higher colloids volume fraction ($\rho_2 = 50 \%$) compensates for the presence of colloids in the cylindrical domains. 

The study of the phase transition induced by the change in NP's volume fraction is in accordance with the phase diagram obtained by Lee \textit{et al} \cite{Lee2002} using SCFT/DFT. Our model allows us also to follow the kinetic relaxation to equilibrium.

\subsection{Colloid Radius Influence on NP Placement }
The effect of the colloidal size in the particle position in the lamella domain has been studied both experimentally \cite{Bockstaller,Okumura_2000} and theoretically \cite{Thompson2001} . It has been found that large particles - relative to the lamella domain- segregate to the lamella center while smaller particles are found preferentially in the lamella interface. This is due to the cost of the copolymer brush to encircle larger colloids. For small particles the decrease in their entropic contribution is not  negligible with respect to the particle's translational entropy.  This behavior has been proved computationally by means of a combination of SCFT for the copolymer and DFT for colloids \cite{Thompson2001} and the phase diagram was successfully obtained.

\begin{figure}[h]
\centering
\includegraphics[width=1.0\linewidth]{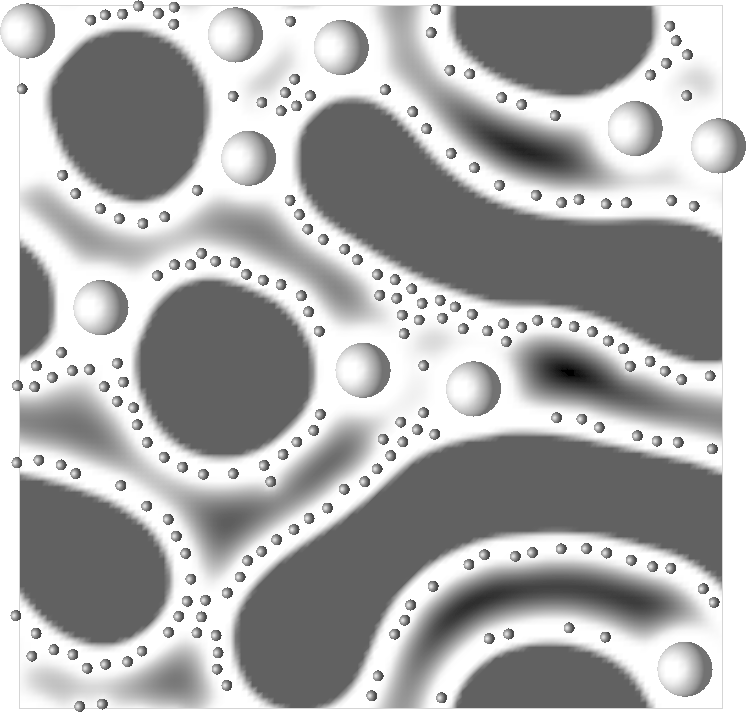}
\caption{$N_s=200$ small ($R_0^s=1.0$) particles and $N_b=10$ big ($R_0^b=6.0$) particles with an assigned affinity $\psici=0.3 $. System size is $128\times 128$. }
\label{fig:bocks.2sized}
\end{figure}

Since our model allows for two different sized kinds of particles we can reproduce the experimental result by Bockstaller \textit{et al}\cite{Bockstaller} by assigning NP's with a weak affinity towards one block. In this regime, the particle size determines the position within the lamella domain. Size-selective segregation occurs if we simulate two kinds of particles with a size ratio   $R_b/R_s = 6.0$ as shown in In Figure \ref{fig:bocks.2sized} where small particles remain mostly  in the interface while big particles are detached towards the center.

\begin{figure}[h]
\centering
\includegraphics[width=\textwidth]{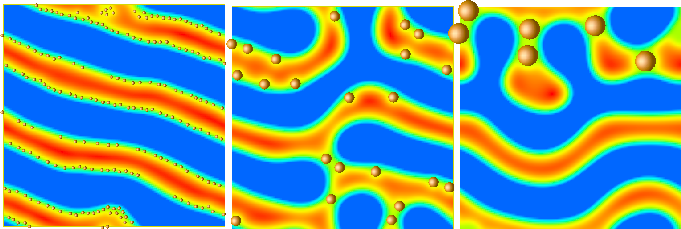}
\caption{Effect of particle's size ($R_0=1.0,3.0,6.0$ from left to right) in its position for a fixed filling fraction and affinity $\psici=0.3$. System size is $128\times 128$ in all cases.}
\label{fig:bocks.r}
\end{figure}

In order to quantify the effect of colloidal size in the particle's position in the lamella domain we can calculate the relative average position inside of the BCP for different values of the hard core radius $R_0$ as shown in Figure \ref{fig:bocks.r}. For a small value of $R_i^0= 1.0$ and $N=200$ (Figure \ref{fig:bocks.r}-left   ) particles simply place themselves in the interface roughly in the same value of the assigned affinity. For a larger value $R_i^0 =3.0$ and while keeping the filling fraction constant, NP's are slighly detached from the interface, as depicted in Figure \ref{fig:bocks.r}-center . For a comparable domain-particle size, particles are completely placed in the center of the domain, as shown in Figure \ref{fig:bocks.r}-right .    

\begin{figure}[h]
\centering
\includegraphics[width=1.0\linewidth]{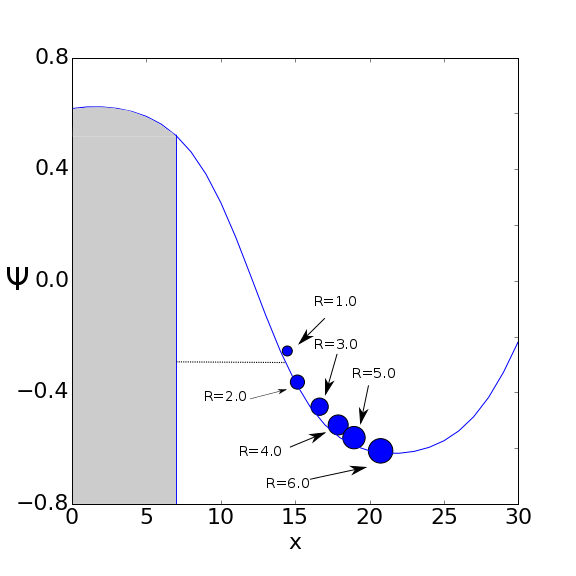}
\caption{Relative average position of nanoparticles in an undistorted lamella profile. The gray area is taken as the reference of center of the domain. The horizontal line shows the affinity that is assigned to all particles. Positive-negative $\psi$ values are inverted in this scheme and particles' sizes are not in scale.}
\label{fig:drift}
\end{figure}

 The results are summarized  in Figure \ref{fig:drift}, where we can observe the drift towards the center of the lamella as we increase the radius of the particles. The small lack of agreement for $R=1.0$ is due to the fact that colloids create two layers in the interface due to a considerably high number of particles. Particles in the second layer contribute as if they had a preference for an slightly out-of-the-interface order parameter value.  Although our model does not take into account the microscopic structure of the BCP, the qualitative effect is properly accounted for  due to the distortion that big particles create in the order parameter. If the distortion is small, the NP can easily fit the interface, but for a larger one the particles need a more homogeneous BCP environment .

\subsection{Hexagonal Packing of Colloids}

So far we have studied the effect of colloids in the the BCP morphology when particles are compatible with one of the copolymers or they are equally soluble in both domains. Contrary to that,  Ploshnik \textit{et al}  \cite{shenhar} performed experiments in the case in which nanoparticles are coated to be incompatible with both blocks, but to different extends, thus triggering a hierarchical structure of the  BCP/colloid composite material and self-assembly of NP into hexagonal packing.

In our CDS/Brownian model we reproduce the incompatibility by assigning an affinity towards one of the copolymers, but larger in magnitude than the maximum value of the order parameter present in the simulation. By doing so we create a distortion in the BCP profile, leading to a penalty in the free energy that can only be minimized by a strong packing of the colloids. 
   
\begin{figure}[h]
\centering
\includegraphics[width=1.1\linewidth]{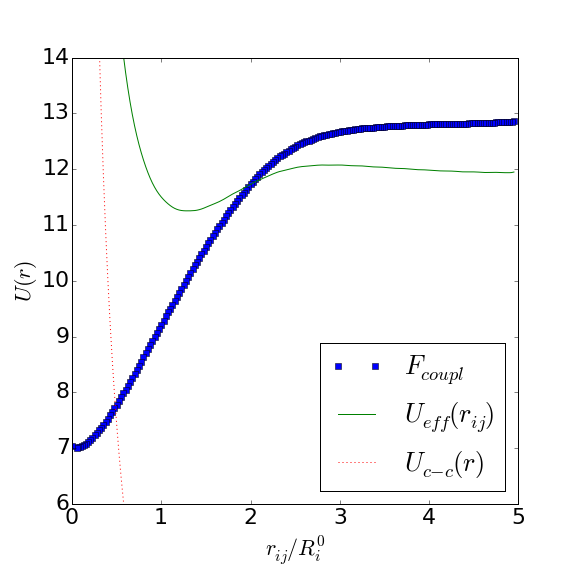}
\caption{Effective potential between particles and colloid-colloid repulsive potential.}
\label{fig:potential}
\end{figure}

\begin{figure}[h]
\centering
\includegraphics[width=1.0\linewidth]{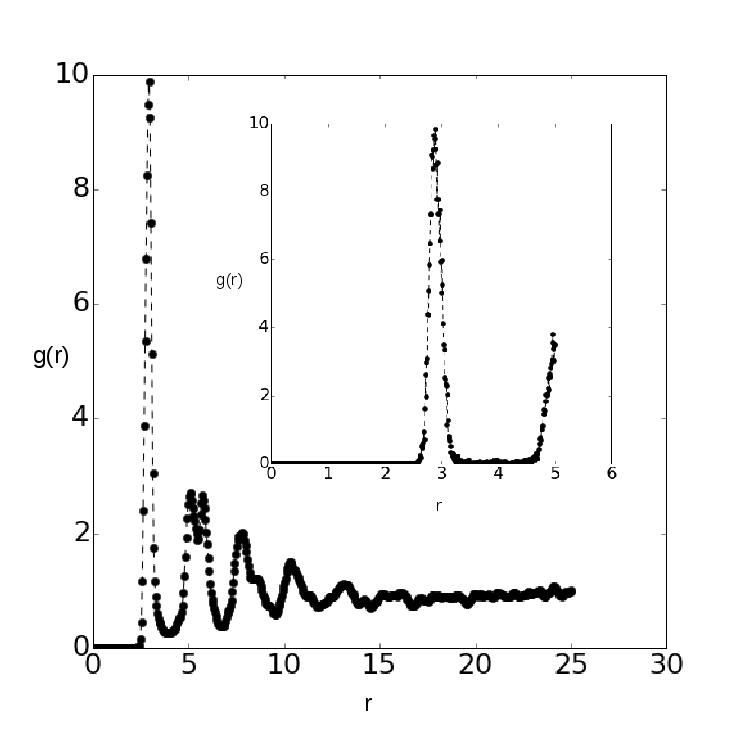}
\caption{Radial distribution function for  Figure \ref{fig:pack.phase_trans} bottom-right. The inset of the first peak is calculated from a subdomain of the system in order to have the right coordination number for the first neighbor, thus removing edge-related effects.  }
\label{fig:gr}    
\end{figure}

\begin{figure}[h]
\centering
\includegraphics[width=1.0\textwidth]{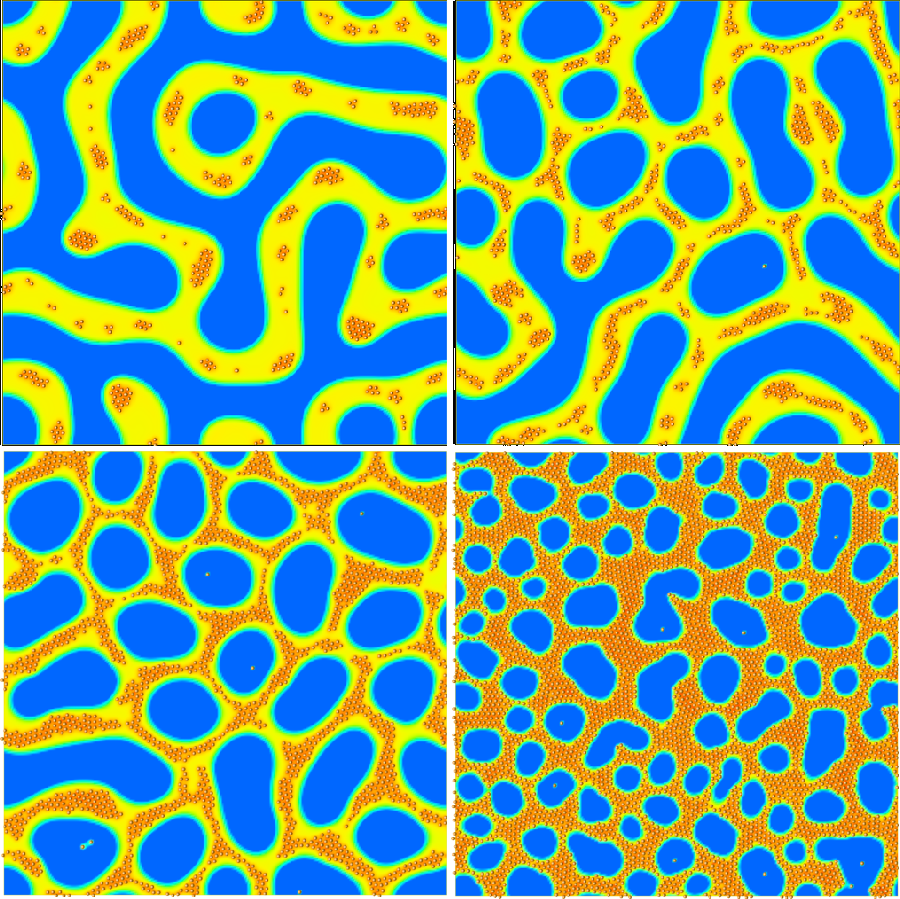}
\caption{Phase transition for a lamella BCP morphology and $N$ NP's with $R_0=1.0$ with increasing volume fraction $\rho = 5  \% ,10 \%,20\%,50\% $ from top-left to bottom-right.
Affinity is $\psici=1.2> \text{max} (|\psi|)$, larger value that the order parameter takes in the absence of   colloids }
\label{fig:pack.phase_trans}
\end{figure}

We find that the overlapping of the distortion of two colloids creates an effective potential as in Figure \ref{fig:potential}. The clear minimum we observe leads to a Lennard-Jones-like potential that in result creates a hexagonal packing, as we can see in Figure \ref{fig:pack.phase_trans}.

Initially we can simulate a system for a small number of NP's at a volume fraction $\rho = 5 \% $,   for which the BCP lamella morphology is not significantly affected, as shown in Figure \ref{fig:pack.phase_trans} top-left.The hexagonal packing already occurs at this low concentration by the formation of clusters of NP's,which resembles particularly correctly the experimental STEM image in Figure  2-I in Ploshnik \textit{et al} work\cite{shenhar} . In this case, we use $\gamma= 1.0$ while in the next simulations we have used $\gamma= 10.0$ because a lower density of particles needs for considerably longer simulation times for particles to create clusters.  For a larger number of particles ($\rho= 10\%$, Figure \ref{fig:pack.phase_trans} top-right) the BCP morphology is distorted leading to a coexistence lamella/cylinder because of the presence of elongated clusters of hexagonally packed colloids. For an even higher volume fraction ($\rho= 20 \%$,Figure \ref{fig:pack.phase_trans} bottom-left) the phase transition is almost complete and BCP domains are distorted into irregular disks. Increasing the NP concentration  leads to an almost complete filling of colloids into their least unfavorable phase, while the drop-like domains are considerably shrinked, as shown in  $\rho= 50\%,$, Figure \ref{fig:pack.phase_trans} bottom-right. 

This phase transition is fundamentally different from the one observed in Figure  \ref{fig:l2c} since the resulting morphology now is disordered instead of hexagonally cylindrical, which is in accordance with experimental results \cite{shenhar} as phase transition from lamella to disordered occurs. Additionally,  it is important to note that the hexagonal packing observed here is  due to a different nature than the one appearing in the previous sections (see Figure \ref{fig:c2l}  ) where packing was a result exclusively of the local high concentration limit (whether it is due to total high concentration value or a local constraint that forces close packing). In this case, hexagonal packing is present even for low concentration of colloids (see Figure \ref{fig:pack.phase_trans} top-left) since colloids minimize the coupling free energy forming clusters.

Using the results from the $\rho = 50 \% $ simulation (Figure \ref{fig:pack.phase_trans} bottom-right) we can calculate the radial distribution function (Figure \ref{fig:gr}) as well as the NP coordination number. The main curve represents $g(r)$ using the whole system which is correct qualitatively. The number of particles associated with the first peak does not result in the expected coordination value $N=6.0$ due to the boundaries created by the inaccessible BCP domains. If we select a small subsystem in which colloids are homogeneously distributed and $g(r)$ is computed again (as shown in the inset in Figure \ref{fig:gr} ) we obtain a value $N=5.97$ for the coordination number after the integration over the first peak as in $N= 2\pi \rho_0 \int_0^{r_0} dr\ r\ g(r)$  which acts as a further proof of the hexagonal packing of NP's.

\begin{figure}[h]
	\centering
	\includegraphics[width=1.0\linewidth]{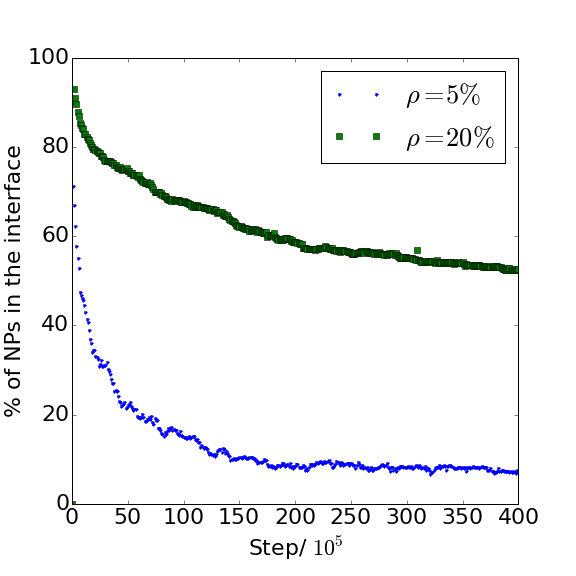}
	\caption{Fraction of NP's that lie in the interface as a function of time.}
	\label{fig:pack.inter}
	\end{figure}

Additional corroboration with the experiments is obtained if we calculate the fraction of colloids present in the interface with respect of the total number. As in the experiment in Figure 8 c in in Ploshnik \textit{et al}  \cite{shenhar} , the fraction of NPs in the interface decays with time, but it does it slower as the filling fraction is increased. In Figure \ref{fig:pack.inter} we can see the dependence for $\rho= 5 \% $ and  $ 20\%$ fulfills the qualitative behavior of the experimental work. We consider interface values those than fulfill $\psi \in [-0.4,0.4] $. This broad definition of the region is necessary since in our model NP's have a very strong affinity with the center of the  domain, resulting in very high penalty for particles outside that region. Nonetheless, Figure \ref{fig:pack.inter} supports the notion proposed in the experimental work \cite{shenhar} that  NP's create clusters in a slower timescale than BCP's self assembly. 

\section{Conclusions}

Cell Dynamics Simulations along with Brownian motion dynamics have been used to simulate the governing equation of a mixture of diblock copolymer and colloids. Our model has proved to be successful when capturing the self assembly of colloids in a BCP matrix and reproducing the phase transition that the morphology of the BCP undergoes as we increase colloid's concentration. The relatively fast nature of the CDS scheme has proved to be crucial to obtain the time evolution of large systems over long times.

The inclusion of two-sized particles has allowed us to reproduce experiments\cite{Bockstaller,Okumura_2000} in which the size of the particle is the main property determining the position of colloids inside of a BCP domain. Hence, our model is able to qualitatively reproduce size-related effects.

Further in our attempt to check the validity of our model, we reproduced the experimental setup in which colloids are incompatible with both blocks in the BCP provided that the affinity is not symmetric towards the blocks. We successfully achieved a clear hexagonal packing, reproducing properties (phase transition and fraction of NP's in the interface) that were present in the experiment\cite{shenhar} . 

In summary, we have presented a model that is able to simulate the effects of size, concentration and chemical properties of NP's immersed in a BCP melt. Order-to-order phase transitions have  been observed, depending on the compatibility of the NP's with each block and NP's volume fraction. 

\hrulefill


Keywords: composites; copolymer/colloid hybrid materials; diblock copolymers; cell dynamics simulations; simulations.

\section*{Text for the Table of Contents}
Cell Dynamic Simulation combined with Brownian Motion are used to describe a mixture of block copolymer and nanoparticles, respectively. We studied the changes in the morphology of the matrix as we include colloids, as well as its assembly depending on their size and affinity with one of the blocks. Several experiments were reproduced.

\clearpage
\bibliography{intro}

\end{document}